\documentclass[twoside,reqno]{lt7-proc}
\usepackage{epsfig,cite}
\usepackage{amssymb,amsmath}
\usepackage{times}
\begin{document}
\sloppy \raggedbottom
\setcounter{page}{1}

%%%%%%%%%%%%%%%%%%%%%%%%%

%
\def\eq#1{(\ref{#1})}
\newcommand{\AAA}{\mathcal A}
\newcommand\CCC{\mathcal{C}}
\newcommand\C{\mathbb{C}}
\newcommand\DDD{{D}}
\newcommand{\HHH}{{\mathcal{H}}}
\renewcommand{\H}{{\mathbb{H}}}
\newcommand{\KKK}{{\mathcal{K}}}
\newcommand\NNN{\mathcal{N}}
\newcommand\N{\mathbb{N}}
\newcommand\R{\mathbb{R}}
\newcommand{\SSS}{{\mathcal{S}}}
\newcommand{\sone}{{\mathbb{S}}}
\newcommand{\WWW}{{\mathcal{W}}}
\newcommand\Z{\mathbb{Z}}
\newcommand{\fs}{\mathfrak{s}}
\newcommand{\fu}{\mathfrak{u}}
\def \bra{\langle}
\def \ket{\rangle}
\def \ie{{\it i.e.}~}
\def \cf{{\it c.f.}~}
\newcommand{\dvol}{\mathrm{dvol}}
\newcommand{\Res}{\mathrm{Res}}
\newcommand{\Spec}{\mathrm{Spec}}
\newcommand{\tr}{\mathrm{tr}}
\newcommand\beq{\begin{equation}}
\newcommand\eeq{\end{equation}}
\newcommand\beqn{\begin{eqnarray}}
\newcommand\eeqn{\end{eqnarray}}
\newcommand\beqa{\begin{eqnarray}}
\newcommand\eeqa{\end{eqnarray}}
\def\vir{{\cal V}{\rm ir}}
\newcommand{\ol}[1]{\overline{#1}}
\def\fa{\forall\,}
\newcommand{\wt}{\widetilde}
\newcommand\req[1]{(\ref{#1})}
\newcommand\mb[1]{\mbox{\rm#1}}
\def\el#1{\label{#1}}
\renewcommand\epsilon{\varepsilon}
\renewcommand\phi{\varphi}
\newcommand\End{\mb{End}}
\newcommand\id{\mb{id}}
\def\polhk#1{\setbox0=\hbox{#1}{\ooalign{\hidewidth
  \lower1.5ex\hbox{`}\hidewidth\crcr\unhbox0}}}
\newcommand\sq{\hbox{\rlap{$\sqcap$}$\sqcup$}}
\newcommand\oldqed{\ifmmode\sq\else{\unskip\nobreak\hfil
\penalty50\hskip1em\null\nobreak\hfil\sq
\parfillskip=0pt\finalhyphendemerits=0\endgraf}\fi}
%
%%%%%%%%%%%%%%%%%%%%%%%%%%

\newpage
\setcounter{figure}{0}
\setcounter{equation}{0}
\setcounter{footnote}{0}
\setcounter{table}{0}
\setcounter{section}{0}

%%%%%%%%%%%%%%%%%%%%%%%%%%

\title{Decoding the geometry of conformal field theories}
\runningheads{ROGGENKAMP, WENDLAND}{DECODING the GEOMETRY of CFT}

%%%%%%%%%%%%%%%%%%%%%%%%%

\begin{start}

\author{\makebox{Daniel Roggenkamp}}{1},
\author{Katrin Wendland}{2}
\address{Department of Physics and Astronomy,         
                   Rutgers, The State University of New Jersey, 
                   136 Frelinghuysen Road,                      
                   Piscataway, NJ 08854-8019 USA}{1}
\address{Lehrstuhl f\"ur Analysis und Geometrie, Institut f\"ur Mathematik, 
Universit\"at Augsburg, D-86135 Augsburg, Germany}{2}

%%%%%%%%%%%%%%%%%%%%%%%%%%

\begin{Abstract}
To certain geometries, string theory associates conformal field theories. 
We discuss techniques to perform the reverse procedure: To recover geometrical data 
from abstractly defined conformal field theories. This is done by introducing
appropriate notions of limits of conformal field theories and their degenerations, 
and by applying techniques from noncommutative geometry.

This note is a 
summary of our work \cite{rowe03}, 
aimed to be less technical than the original paper, along with some new calculations
confirming our interpretation of the rescaled limiting zero mode 
of the Virasoro field.
\end{Abstract}
\end{start}

%%%%%%%%%%%%%%%%%%%%%%%%
%
%
\section{Introduction}
In string theory, it is believed that one can associate quantum field 
theories to certain geometries by a so-called non-linear sigma model construction.
The corresponding conformal quantum field theories (CFTs) naturally 
exhibit traces of the original ``target space geometry". 
This connection between geometry and CFT is a continuous 
driving force both in geometry and in string theory, as is exemplified by the
success stories of mirror symmetry and T-duality, to name just two examples.
However, in general, quantisation requires 
renormalisation and thus obscures the underlying geometry in non-linear sigma models.
It is therefore difficult
to describe how exactly geometry is encoded in CFT. 
It is nevertheless appealing to investigate this problem, since CFTs 
can be abstractly defined in more general contexts. Moreover, the 
applications should be manifold, by the above,
for example, in the study of mirror symmetry and T-duality. 

An exact correspondence between CFT and geometry is expected in 
general only in so-called large volume regions of the target space, where quantum
effects become small. More precisely, the Riemannian metric of the underlying target
space needs to be considered close to a degenerating metric.
On the CFT side one should therefore expect an analogous 
degeneration phenomenon. In \cite{rowe03} we introduce a notion of limiting
processes for CFTs which we argue leads to precisely such
phenomena. Moreover, we show that the limiting data can be used to associate geometric
interpretations to degenerate limits of CFTs, and we substantiate
our claim by proving that the procedure works for the large level 
limit of the bosonic unitary Virasoro minimal models of A-type. To achieve this, we employ
ideas of Fr\"ohlich and Gaw{\polhk{e}}dzki \cite{frga93}, using the fact that the 
energy operator in a semi-classical limit of CFTs should yield
a generalised Laplacian on the target space. While its spectrum belongs to the 
data recovered in the semi-classical limit, in general, one cannot hear the shape
of a drum. However, Fr\"ohlich and Gaw{\polhk{e}}dzki argue that the operator
product expansion (OPE) of CFT 
yields the necessary hearing aid: It promotes a subspace of the Hilbert
space (interpreted as space of square integrable functions on the target) to an
algebra of (continuous) functions. Altogether, one obtains a triple of data which is 
closely related to Connes' spectral triples \cite{co85,co90,co96}. In particular,
one can read off a topological space as target space together with two conformally
equivalent Riemannian metrics on it, one the target space metric and the other the
dilaton-corrected metric. 

The present paper serves two main purposes: On the one hand, we wish to give a more
digestible summary of our previous work \cite{rowe03}, by focussing on the central
ideas instead of technical details. On the other hand, thanks to comments by Maxim
Kontsevich, we have meanwhile obtained additional evidence for the validity of our 
geometric interpretation of limiting processes in CFTs, which we
present here. 

The structure of the paper is as follows:
In section \ref{defs}, we recall the main ideas behind our definition of limiting sequences
of CFTs. The section contains a brief review of our abstract definition
of such theories, since it is fundamental to the notion of limits.
Section \ref{interpret} begins with a review of those properties of the limiting data
which allow us to associate geometric interpretations to them. We then recall the
relevant techniques from noncommutative geometry. Finally, we state the so-called
seven-term identity, and we prove that 
this identity can be viewed as an analogue of the Leibniz rule, characterising 
differential operators of order two without constant term and with symmetric symbol.
The seven-term identity is also the central theme of section \ref{seven}. Here we prove
that in fact the operator $H^\infty$, obtained by rescaling the energy operator 
in a degenerate limit of CFTs, obeys the seven-term identity.
In other words, our interpretation of $H^\infty$ as generalised Laplacian without
constant term is confirmed. We end this note with some conclusions and an outlook
in section \ref{conc}.
\section{Limits of CFTs: The definition}\el{defs}
The aim of this work is to find a general procedure to extract geometry from 
conformal field theory (CFT), using intrinsic CFT data only. Thus we need
to work with an abstract definition of CFT. Here, we restrict ourselves to
stating the main ingredients of such a definition for
the cases of interest to us. Further 
details can be found in \cite{rowe03} and in the references given there.

A unitary two-dimensional Euclidean conformal field theory (in short a CFT), roughly, 
is a unitary representation of two commuting copies $\vir_c$, $\ol{\vir}_{\ol c}$ 
of the Virasoro algebra on a complex vector space
$\HHH$ with positive definite scalar product $\langle\cdot,\cdot\rangle$ 
and complex conjugation $\ast\colon\HHH\rightarrow\HHH$. It is 
assumed that the commuting self-adjoint operators $L_0,\,\ol L_0$ given by
the zero modes of the left and the right handed Virasoro algebra, respectively,
are simultaneously diagonalisable such that 
$$
\HHH=\bigoplus_{(h,\ol h)\in R}\HHH_{h,\ol h},\quad R\subset\R^2,
$$
where all eigenvalues $h,\,\ol h$ 
are non-negative,
all simultaneous eigenspaces $\HHH_{h,\ol h}$ of $L_0,\ol L_0$
are finite dimensional, and such that
$\HHH_{0,0}=\ker(L_0)\cap\ker(\ol L_0)$ is one-dimensional. 
A unit generator 
$\Omega\in\HHH_{0,0}$ is chosen, called the vacuum.
We introduce the ``graded dual'' 
$$
\check\HHH^\ast:=\bigoplus_{(h,\ol h)\in R}(\HHH_{h,\ol h})^\ast
$$ of $\HHH$ 
as well 
as the natural map  
$\HHH\longrightarrow \HHH^\ast, \,\psi\mapsto\psi^\ast\in\check\HHH^\ast$
such that for every $\chi\in\HHH$ one has $\psi^\ast(\chi)=\langle\psi,\chi\rangle$.

In addition to this 
structure, the vector space $\HHH$ in a CFT carries so-called
$n$-point functions which in turn encode the OPE. The
$n$-point functions are multilinear, symmetric maps from 
$\HHH^{\otimes n}$ into functions in $n$ (holomorphic) complex coordinates
$z_1,\ldots,z_n$ on 
a Riemann surface, which are real analytic 
outside the partial diagonals and with prescribed pole behaviour near the partial
diagonals. These $n$-point functions 
``transform covariantly under infinitesimal conformal
transformations'', a notion which can be made precise in terms of the representation
theory of $\vir_c\oplus \ol\vir_{\ol c}$. In particular, insertion of the vacuum into the 
$i^{\scriptsize\rm th}$ component of an $n$-point function yields a function which 
is independent of $z_i$ and which can be viewed as an $(n-1)$-point 
function. There are various additional
consistency conditions, the most important of which, from our point of view, ensures
that all $n$-point functions can be recovered from the structure constants of the 
three-point functions, that is from certain maps 
$$
C\colon \check\HHH^\ast\otimes\HHH\otimes\HHH\longrightarrow\C
$$ 
which obey
\beq\el{psistar}
\fa \psi,\chi\in\HHH:\quad\quad
C(\psi^\ast,\Omega,\chi)=\langle\psi,\chi\rangle.
\eeq
The expert will notice that $C$ is related to the usual three-point functions
by 
$$
C(\psi^\ast,\varphi,\chi) = \langle \psi^\dagger(\ol x^{-1})\,\varphi(z)\,\chi(w)\rangle_{\mid x=w=0, z=1}.
$$
In fact, the $n$-point functions can be recovered from the restriction of
$C$ to the space of all highest weight vectors $\HHH^\WWW$ (and its graded dual)
with respect to an appropriate choice of W-algebra $\WWW$ such that 
$\vir_c\oplus\ol\vir_{\ol c}\subset\WWW\subset\ker (L_0)\oplus\ker(\ol L_0)$. 
The elements of $\HHH^\WWW$ are the so-called 
primaries with respect to $\WWW$.
In what follows, it will be more convenient to encode $C$
in the following way:
%terms of a representation of $\HHH^\WWW$ on itself as follows:
For every $\phi\in\HHH^\WWW$, we define $A_\phi\in\End(\HHH^\WWW)$ such that
$$
\fa \psi,\chi\in\HHH^\WWW: \quad
C(\psi^\ast,\phi,\chi) = \langle \psi, A_\phi(\chi)\rangle.
$$
The axioms of CFT ensure that this weak definition of $A_\phi$
indeed gives a well-defined linear operator $A_\phi$
for every $\phi\in\HHH^\WWW$. Due to \req{psistar} we immediately find
$A_\Omega=\id$. The expression $A_\phi(\chi)$ can be viewed
as a truncation  to $\HHH^\WWW$ of the OPE between $\phi$ and $\chi$, where
$\phi, \chi\in\HHH^\WWW$. This 
implies $A_\phi\circ A_\chi \neq A_{A_\phi(\chi)}$ in general, yielding
the algebra $\AAA\subset\End(\HHH^\WWW)$ which is formally generated by all 
operators
$A_\phi$ with $\phi\in\HHH^\WWW$ rather complicated.
In fact, $A_\phi\circ A_\chi$
may not be of the form $A_\psi$ for any $\psi\in\HHH^\WWW$.

In what follows, of all these structures we will mainly use:
\begin{itemize}
\item
The W-algebra $\WWW\subset\ker L_0\oplus\ker\ol L_0$ with
$\WWW\supset\vir_c\oplus\ol\vir_{\ol c}$, and its associated
space of primaries $\HHH^\WWW$.
\item
The action of the so-called energy operator $H:=L_0+\ol L_0$ on $\HHH^\WWW$.
\item
The algebra $\AAA\subset\End(\HHH^\WWW)$ generated by the operators $A_\phi$ with $\phi\in\HHH^\WWW$, 
where the map $\phi\mapsto A_\phi$ is
obtained from $C$ as above, and $C$ encodes the $n$-point functions.
\end{itemize}
For later convenience, we list the resulting structures in the case
of the CFT which describes a free boson compactified on a circle of radius 
$R\in\R^+$:
\begin{itemize}
\item
$\WWW=\fu(1)\oplus\ol{\fu(1)}$, and the space 
$\HHH^\WWW_R$ of corresponding primaries has a basis 
$\left\{ | m,n\rangle_R ,\; m,n\in\Z\right\}$ indexed by pairs of integers.
\item
The basis of $\HHH^\WWW$ given above is an eigenbasis for the energy operator $H$, namely
$$
H \left( \vphantom{ {1\over2}} | m,n\rangle_R \right)
= {1\over2}\left( {m^2\over R^2} + n^2 R^2 \right) | m,n\rangle_R.
$$
\item
The algebra $\AAA\subset\End(\HHH^\WWW)$ is
generated by the operators $A_{| m,n\rangle_R}$, $m,n\in\Z$, with 
%\beq\el{circlealgebra}
$$
A_{| m,n\rangle_R} \left( | m^\prime,n^\prime\rangle_R \right)
= (-1)^{mn^\prime} | m+m^\prime,n+n^\prime\rangle_R.
$$
\end{itemize}
Indeed, the above CFTs form a family parametrised by $R\in\R^+$, 
and it is natural to study their behaviour 
when $R$ converges to $0$ or $\infty$. 
To be able to address such problems
a notion of convergent sequences of CFTs is needed. 
The limit of a convergent sequence
of CFTs should exhibit those properties which are common to 
almost all of its members. To extract such common properties, we use the 
notion of direct limits of vector spaces, and we impose conditions on the sequence
of CFTs which allow us to define some of the usual structures of a CFT on such
a direct limit. It is crucial, however, that a limit of a sequence of CFTs according
to our definitions need not be a full-fledged CFT itself. 

In more detail, we define a sequence of CFTs to consist of a family of CFTs whose underlying
vector spaces $\left\{\HHH^i\right\}_{i\in I}$  form a direct system.
In other words, $I$ is an ordered index set, and
vector space homomorphisms $f_i^j\colon \HHH^i\longrightarrow\HHH^j$ are given
for all $i\leq j$ such that $f_i^i=\id,\, 
f_j^k\circ f_i^j=f_i^k$ for $i\leq j\leq k$.
We additionally assume that $f_i^j\ast=\ast f_i^j$ and
that the $f_i^j$   map the most fundamental states in our CFTs to 
one another, that is the vacua and the states created by the holomorphic and the 
antiholomorphic Virasoro fields. 
These conditions suffice to define a vector space $\KKK^\infty$ obtained as direct limit 
$\KKK^\infty=\lim\limits_{\longrightarrow} \HHH^i$. Elements of $\KKK^\infty$ hence are 
equivalence classes of elements of the $\HHH^i$, whose representatives form
sequences $\{f_i^j(\psi_i)\}_{j\geq i}$ with $\psi_i\in\HHH^i$.

In \cite{rowe03} we give a detailed definition for the notion of convergence
of a sequence of CFTs with underlying vector spaces $\wt\HHH^i$. 
This definition is quite technical, such that here we prefer to focus on a summary.
Convergence roughly 
means that 
the direct system $\wt\HHH^i$ has a direct subsystem $\HHH^i\subset\wt\HHH^i$ such
that after restriction to the $\HHH^i$ one has:
\begin{enumerate}
\item 
For all $\psi,\phi,\chi,\rho\in\HHH^i$,
the four-point function
$$
z\longmapsto\langle f_i^j(\psi)\mid f_i^j(\phi)(1)\,f_i^j(\chi)(z) \mid f_i^j(\rho)\rangle
$$
converges for $j\rightarrow\infty$ as a real analytic function 
and with the standard behaviour near the singularities $z\in\{0,1,\infty\}$.
\item
The maps $f_i^j$ respect a collection of chosen W-algebras 
$\WWW^i\subset\ker(L_0^i)\oplus\ker(\ol L_0^i)$ in the following sense: 
The $f_i^j$ respect an appropriate decomposition of the $\HHH^i$ into common
$L_0^i$ and $\ol L_0^i$ eigenspaces,
primaries with respect to $\WWW^i$ are mapped to primaries with respect 
to $\WWW^j$, and none of these primaries represents zero in $\KKK^\infty$. 
\item
The convergence of three-point functions is uniform in an appropriate sense,
as is the convergence of four-point functions involving primaries. Moreover, 
only finitely many homogeneous components $\phi^i\in\HHH^i_{h_\phi^i, \ol h_\phi^i}$ 
of each truncated OPE 
represent elements of $\KKK^\infty$ with 
$\lim\limits_{i\rightarrow\infty} C^i((\phi^i)^\ast,\Omega^i,\phi^i)\neq 0$
and $\lim\limits_{i\rightarrow\infty} \left(h_\phi^i+ \ol h_\phi^i\right)= 0$.
\end{enumerate} 
In \cite{rowe03} we show that our definition of limits of CFTs allows us to obtain the
following structures from a convergent sequence of CFTs: 

A unique vector
$\Omega\in\KKK^\infty$ exists which is represented by the vacua in the $\HHH^i$. 
There is a direct limit $(\check\KKK^\infty)^\ast$ of the $(\check\HHH^i)^\ast$
and an induced isomorphism $\psi\mapsto\psi^\ast$ between $\KKK^\infty$ and
$(\check\KKK^\infty)^\ast$. The limiting structure constants $C^\infty$ 
of three-point functions are well-defined on $\KKK^\infty$, and they induce a sesqui-linear form 
$\langle\psi,\chi\rangle :=C^\infty(\psi^\ast,\Omega,\chi)$ on $\KKK^\infty$
which however need
not be positive definite. Setting 
$$
\NNN^\infty:=\left\{\nu\in\KKK^\infty\mid C^\infty(\nu^\ast,\Omega,\nu)=0\right\},
$$
our conditions on convergence ensure that $C^\infty$ descends to well-defined structure
constants on $\HHH^\infty:=\KKK^\infty/\NNN^\infty$, where $\langle\cdot,\cdot\rangle$
is positive definite. On $\HHH^\infty$ there is a well-defined action of a limiting
W-algebra $\WWW$ and thus in particular of a limiting Virasoro algebra with zero modes
$L_0,\,\ol L_0$, as well as a notion of primaries. In fact, $\HHH^\infty$
decomposes into a direct sum of  simultaneous eigenspaces 
$\HHH^\infty_{h,\ol h}$ of the operators $L_0, \ol L_0$. 
The limiting energy operator is denoted by $H:=L_0+\ol L_0$.
Furthermore, a truncated OPE $A_\phi(\chi)$ can be defined between any 
two primaries $\phi,\chi$ in $\HHH^\infty$.

We thus in particular obtain the following data from any limit of a convergent
sequence of CFTs, where for simplicity we assume that the elements
of the sequence are indexed by some $I\subset\R$:
\begin{itemize}
\item
A vector space $\H^\infty:=\ker H\subset\HHH^\infty$ with positive definite
scalar product $\langle\cdot,\cdot\rangle$.
\item
A limiting energy operator $H^\infty$ on $\H^\infty$ obtained by rescaling $H$ appropriately:
If there is a function $r:I\rightarrow\R$ with $\lim\limits_{\longrightarrow} r=0$ 
such that for all $\varphi\in\H^\infty$ 
represented by homogeneous vectors $(\varphi^i\in\HHH_{h_i,\ol h_i}^i)_i$ the rescaled energies
${1\over r(i)}(h_i+\ol h_i)$ remain finite in the limit, 
then define $H^\infty\varphi=E\varphi$ 
with $E:=\lim\limits_{\rightarrow}{1\over r(i)}(h_i+\ol h_i)$.
\item
A limiting zero mode algebra $\AAA^\infty\subset\End\left((\HHH^\infty)^\WWW\right)$ 
generated by the operators $A_\phi$ with $\phi\in\H^\infty$,
$$
C^\infty(\psi^\ast,\phi,\chi) = \langle \psi, A_\phi(\chi) \rangle
\quad \mb{ for all } \quad  \psi,\chi\in(\HHH^\infty)^\WWW.
$$
\end{itemize}
It is important to keep in mind that the structure found in the limit of a convergent
sequence of CFT severely depends on the choice of the direct system, \ie the maps $f_i^j$. In \cite{rowe03},
we give several instructive examples to this effect. However, in many cases there
are natural choices for the maps $f_i^j$. For instance for the free boson 
on a circle, a natural choice is induced by 
$f_R^{R^\prime}\colon | m,n\rangle_R \mapsto | m,n\rangle_{R^\prime}$
for all $m,n\in\Z$.
It is not hard to check convergence and that then
the following structures arise in the limit $R\rightarrow\infty$:
\begin{itemize}
\item
Each primary $|m,0\rangle_R\in\HHH_R^\WWW$ with $m\in\Z$ represents a primary 
$|m,0\rangle_\infty\in\HHH^\infty$,
and in fact $\H^\infty$ has orthonormal basis $\left\{| m,0\rangle_\infty, m\in\Z\right\} $.
The highest weight states $|m,n\rangle_R$ with $n\neq 0$ and their descendants do not contribute to the limit,
because in the limit $R\rightarrow\infty$ their energies diverge and they decouple from the rest of the theory. 
\item
The energy eigenvalues of representatives of elements in $\H^\infty$ can be rescaled 
by a factor of ${1\over r(R)}=R^2$, so 
$H^\infty \left(| m,0\rangle_\infty\right) = {m^2\over2}| m,0\rangle_\infty$.
\item
The limiting zero mode algebra $\AAA^\infty\subset\End\left((\HHH^\infty)^\WWW\right)$ is
generated by the operators $A_{| m,0\rangle_\infty}$ with $m\in\Z$, and
\beq\el{circlemult}
A_{| m,0\rangle_\infty}(| m^\prime,0\rangle_\infty)=| m+m^\prime,0\rangle_\infty.
\eeq
\end{itemize}
In \cite{rowe03} we also show that the sequence of diagonal bosonic unitary Virasoro minimal models
can be given the structure of a sequence of CFTs, which converges. We will recall the
interpretation of its limit below in somewhat greater detail.
\section{Limits of CFTs: Extracting geometric interpretations}\el{interpret}
The notion of convergence of sequences of CFTs given in the last section
allows for certain degeneration phenomena to occur. We will argue that precisely these
degeneration phenomena allow us to extract geometric data from limits of CFTs. 

As emphasized above, the limit of a convergent sequence of CFTs is not necessarily a 
full-fledged conformal field theory. 
Indeed, our definition of convergence assures the existence of 
limiting correlation functions on the sphere which
satisfy the appropriate conditions, but correlation functions on higher genus surfaces can diverge due to
degenerations of the spectra of $L_0$ and $\ol L_0$.  In particular, it may happen that
the space $\H^\infty$ 
of states whose energy converges to zero, in the limit has infinite dimension.
This is already the case for the sequence of CFTs associated to the free boson on the circle, 
as we have seen above. However, our notion of convergence ensures that this space and its associated
zero mode algebra $\AAA^\infty$ have surprisingly simple properties:

The vanishing of the conformal weights for every $\phi\in\H^\infty$ ensures
\\[3pt]
{\bf Lemma:} \textsl{In a convergent family of CFTs, for every 
$\varphi\in\H^\infty$, the corresponding zero mode
$A_\varphi\in\AAA^\infty$ acts as an endomorphism of each simultaneous eigenspace 
$\HHH^\infty_{h,\overline{h}}$ of $L_0, \ol L_0$. In particular, the zero mode algebra
$\AAA^\infty$ acts on $\H^\infty$.}
\\[3pt]
Moreover, crossing symmetry of four-point functions on the sphere implies 
\\[3pt]
{\bf Proposition:} \textsl{The zero mode algebra $\AAA^\infty$ associated to a convergent 
family of CFTs is commutative, $A_\phi\circ A_\chi=A_{A_\phi(\chi)}$ 
with $A_\phi(\chi)\in\H^\infty$ for all $\phi,\chi\in\H^\infty$.}
\\[3pt]
Therefore, by means of the Gelfand-Naimark theorem to every such degenerate limit of CFTs, 
one can associate a topological space $X=\Spec(\ol{\AAA^\infty})$
with $\ol{\AAA^\infty}$ an appropriate completion of $\AAA^\infty$. In other words,
the zero mode algebra $\AAA^\infty$ 
can be interpreted to generate the algebra of continuous functions on 
a topological space $X$.
%, and multiplication in 
%this algebra is given by pointwise multiplication of functions. 
The limiting Hilbert space $\HHH^\infty$
can be regarded as space of sections $\HHH^\infty=\Gamma(\SSS)$ of a sheaf $\SSS$ of vector spaces on $X$.

For example, for the limit $R\rightarrow\infty$ of the sequence of CFTs describing
a free boson on a circle of radius $R$, we easily identify $|m,0\rangle_\infty$
with the continuous function $x\mapsto\exp(\sqrt{-1}\,m\,x)$ on 
$X=\sone^1=\R^1/(x\sim x+2\pi)$, obeying the multiplication law \req{circlemult}. The sheaf $\SSS$ is a trivial vector bundle
on $X$ with fibers given by the vacuum representation of $\fu(1)\oplus\ol{\fu(1)}$.

In general the structure of $\SSS$ is more interesting. For the $\Z_2$-orbifold of the free boson for instance, 
the twisted sectors contribute to $\SSS$ as sections of skyscraper sheaves over the orbifold 
fixed points, \cf \cite{rowe03}. For brevity we will refrain from discussing the structure of $\SSS$ any further, instead we will  argue how the limiting data 
$(\H^\infty, H^\infty, \AAA^\infty)$ obtained above can yield a geometric
interpretation on a space $X$ equipped with a Riemannian metric (up to scaling). Here we
use ideas from Connes' noncommutative geometry \cite{co85,co90,co96}, following
\cite{frga93,koso00}. Namely, apart from the identification of $\ol{\AAA^\infty}$
as the algebra of continuous functions on $X$, 
the vector space $\H^\infty$ should be interpreted
as a space of square integrable functions on $X$,
where $H^\infty$ is densely defined and
acts as generalised Laplacian. This will put
$(\H^\infty, H^\infty, \AAA^\infty)$ into the context of Connes' spectral triples.
In more detail, in favourable cases there exist conformally equivalent
Riemannian metrics $g$ and $\wt g$ on $X$ such that
$\H^\infty$ generates $L^2(X,\dvol_g)$,
and $H^\infty$ is a generalised Laplacian associated to 
$\wt g$. Writing
$\dvol_g = e^{2\Phi} \dvol_{\wt g}$ with $\Phi\in C^\infty(X)$,  
in coordinates on $X$, the linear operator $2H^\infty$ corresponds to 
the generalised Laplacian
\beq\el{localform}
\Delta_{\wt g}^\Phi =-e^{-2\Phi}\sqrt{\det(\wt g^{-1})}
\sum_{i,j}\partial_i\,e^{2\Phi}\sqrt{\det(\wt g)}\wt g^{ij}\partial_j\,.
\eeq
The function $\Phi$ is called a dilaton field.

For the infinite radius limit of CFTs describing a free boson on 
a circle we indeed easily identify $2H^\infty$ with the ordinary Laplacian
$-{d^2\over dx^2}$, such that $g$ is the flat metric on $\sone^1$ and the 
dilaton $\Phi=0$ is trivial. In general, the ideas described above are in 
accord with the aim to find a reversal of non-linear sigma model constructions.
As is explained in more detail in \cite{frga93,rowe03}, the ansatz
$(\ol{\H^\infty}, 2H^\infty, \ol{\AAA^\infty})=(L^2(X,\dvol_g),\Delta_{\wt g}^\Phi, C^0(X))$
gives a mathematical formulation of a semiclassical limit. The degeneration
of CFTs yields a degenerating family of operators $r(i)2H^\infty$, which gives rise to families of 
degenerating metrics ${g\over r(i)}, {\wt{g}\over r(i)}$. The operator $2H^\infty$ and the metrics $g,\wt{g}$
are only defined up to an overall scale and serve as a reference for the degeneration. For the infinite radius limit
of the free boson, for instance, one obtains the degenerating family of 
(generalised) Laplacians
 $-{1\over R^2}{d^2\over dx^2}$ associated to the family of flat 
metrics on the circle of radius $R$, as expected.

It is not hard to explicitly recover the Riemannian metric $g$ as well as the 
dilaton $\Phi$ from 
$(\ol{\H^\infty}, 2H^\infty, \ol{\AAA^\infty})
=(L^2(X,\dvol_g),\Delta_{\wt g}^\Phi, C^0(X))$,
once $X$ is known, \cf \cite{frga93}. In \cite{rowe03} we collect
further evidence that this procedure works in many cases. In particular, 
we show that it is valid for an example which does not arise from
a non-linear sigma model construction by proving
\\[3pt]
{\bf Proposition:} \textsl{The family of bosonic unitary Virasoro minimal models
of A-type gives a convergent sequence of conformal field theories, 
whose large level limit has geometric interpretation on the unit interval
$X=[0,\pi]$ with flat metric $g\equiv1$ and dilaton $\Phi$ such that
$\exp(2\Phi(x))={2\over\pi} sin^2(x)$ for $x\in X$.}
\\[3pt]
This result is in accord with the predictions in \cite[\S3.3]{frga93}.
Namely, one argues that the Virasoro minimal model at level $k\in\N$
is given by a coset model
$$
\left(\fs\fu(2)_{k-1}\oplus \fs\fu(2)_1\right) \slash \fs\fu(2)_k.
$$
This implies that as $k\rightarrow\infty$, one approximately has an
$\fs\fu(2)$ WZW model at level $1$ which is gauged by the adjoint action of $\fs\fu(2)$.
The gauge projection is elegantly implemented by 
$\fs\fu(2)\rightarrow U(1)/\Z_2$, $h\mapsto\tr(h)$ with
$U(1)/\Z_2\cong [0,\pi]$ in accord with our proposition. 
By \cite[chapter 4.2]{ts93}, the dilaton $\exp{(2\Phi)}$
can be obtained from the Lagrangian of a gauged WZW model 
as the coefficient in front of those contributions which are quadratic in the
gauge field. Such contributions have been determined \cite[(2.3)]{gaku89},
and a straightforward calculation confirms $\exp{(2\Phi(x))}\sim sin^2(x)$.

After completion of \cite{rowe03},
Maxim Kontsevich pointed out to us that according to our claim, 
the rescaled energy operator $H^\infty$
on $X$ corresponds to a second order differential operator 
without constant term, and that it therefore must satisfy the following 
so-called seven-term identity:
\\[3pt]
{\bf Proposition:} \textsl{For $a,b,c\in\AAA^\infty$, where 
by misuse of notation we identify elements $a\in\AAA$ canonically with
$a\cdot\Omega\in\H^\infty$}\mb{:}
\begin{eqnarray}\label{7term}
H^\infty(abc)-H^\infty(ab)c-aH^\infty(bc)-bH^\infty(ca)&&\nonumber\\
+H^\infty(a)bc+aH^\infty(b)c+abH^\infty(c)%
&=&0\,.
\end{eqnarray}
Vice versa, \req{7term} can be viewed as an analog of the Leibniz rule 
for second order differential operators without constant term. Recall
that a linear operator $\DDD$ acting on functions $f\colon X\rightarrow\C$
on a Riemannian manifold $X$ is called a derivation if $\DDD$ obeys the
Leibniz rule
$$
\fa f,h:\quad\quad
\DDD(fh)=f\,\DDD(h)+h\,\DDD(f).
$$
A linear operator acts as a derivation on smooth
functions on $X$ if and only if, in local coordinates on $X$,
it is a differential
operator of first order without constant term. 
There is an analogous statement for differential
operators of order two:
\\[3pt]
{\bf Proposition:} \textsl{Consider a  Riemannian manifold $(X,g)$ 
and a linear operator $H$ which is densely defined on
$L^2(X,\dvol_g)$. Then $H$
obeys the seven-term identity \mb{\req{7term}} if and only if $H$ is a 
second order differential operator without constant term and with
symmetric symbol, \ie}
\beq\el{sop}
H=\sum_i \smash{{1\over2}} \left( \DDD_i\circ\wt\DDD_i + \wt\DDD_i\circ\DDD_i\right)
+ \DDD
\eeq
\textsl{for derivations $\DDD_i,\,\wt\DDD_i,\,\DDD$.}\\[3pt]
{\bf Proof:}
That $H$ as in \req{sop}
with derivations $\DDD_i,\,\wt\DDD_i,\,\DDD$ obeys  
\req{7term} is an immediate consequence of the Leibniz 
rule for each of the operators $\DDD_i,\,\wt\DDD_i,\,\DDD$. 

Conversely, assume that $H$ obeys \req{7term}. With $\H\subset L^2(X,\dvol_g)$
its domain of definition, we consider the family 
$\left\{ \DDD_f \right\}_{\mid f\in\H}$ of linear operators defined by
$$
\fa f\in\H:\quad\quad
\DDD_f\colon \H\longrightarrow\ol\H,\quad
\DDD_f(h):= H(fh)-f\,H(h)-h\,H(f).
$$
Then the seven-term identity \req{7term} implies the Leibniz rule for each $\DDD_f$.
Since $\DDD_f(h)=\DDD_h(f)$ for all $f,h\in\H$, there are derivations
$\DDD_i,\,\wt\DDD_i$ such that
$$
\fa f,h\in\H:\quad\quad
\DDD_f(h) = \sum_i\left( \DDD_i(f)\,\wt\DDD_i(h) + \wt\DDD_i(f)\,\DDD_i(h) \right).
$$
Now consider the linear operator
$$
\DDD:= H - \sum_i {1\over2} \left( \DDD_i\circ\wt\DDD_i + \wt\DDD_i\circ\DDD_i\right).
$$
By construction, for all $f,h\in\H$ we have:
\begin{eqnarray*}
\DDD(fh) &=& \DDD_f(h) + f\,H(h)+ h\, H(f)\\
&&- \sum_i {1\over2} \DDD_i\left( f\,\wt\DDD_i(h)+h\,\wt\DDD_i(f)\right)
- \sum_i {1\over2} \wt\DDD_i\left( f\,\DDD_i(h)+h\,\DDD_i(f)\right) \\
&=& f\,\DDD(h)+h\,\DDD(f).
\end{eqnarray*}
In other words, $\DDD$ is a derivation, as claimed.
\vspace{3pt}
\oldqed
An operator $H$ as in the above proposition which obeys the seven-term identity and in 
addition is symmetric with respect to the $L^2$-scalar product on $\H$
can be brought into the form \req{localform}.
We will therefore prove that \req{7term} holds in general for a limit
of CFTs, according to our definition. This is done in the following section
and serves as further evidence for the desired interpretation 
$(\ol{\H^\infty}, 2H^\infty, \ol{\AAA^\infty})=(L^2(X,\dvol_g),\Delta_{\wt g}^\Phi, C^0(X))$
of limiting data of convergent sequences of CFTs.
\section{The seven-term identity}\el{seven}
Following remarks of Kontsevich, in this section we show that the
rescaled energy operator $H^\infty$ acting on the algebra
$\AAA^\infty$ in a degenerate limit of conformal field theories obeys
the seven-term identity (\ref{7term}). As argued above, this implies that $H^\infty$
corresponds to a
second order differential operator without constant term which in local coordinates 
can be brought into the form (\ref{localform}).

Consider a one-parameter family of conformal field theories parametrised
by $\epsilon\in (0,1)$, which yields a convergent sequence and
degenerates as $\epsilon\rightarrow0$. We
assume that all correlation functions are analytic in $\epsilon$ at $0$. 
Let $\phi_i^\epsilon\in\HHH^\epsilon_{h_i^\epsilon,\ol h_i^\epsilon}$ represent $\phi_i\in\HHH^\infty$, 
such that the respective eigenvalues $h_i^\epsilon,\bar h_i^\epsilon\rightarrow 0$
for $\epsilon\rightarrow 0$. Consider the four-point functions
on the sphere 
$$
F_{1234}^\epsilon(z,\bar z)
\!:=\!\bra \phi_1^\epsilon(\infty)\phi_2^\epsilon(1)\phi_3^\epsilon(z,\bar
z)\phi_4^\epsilon(0)\ket^\epsilon 
\!=\!\sum_i C_{34}^i(\epsilon)
C_{2i}^1(\epsilon) f_i^\epsilon(z) \bar f_i^\epsilon(\bar z)\,.
$$
Here $C_{ij}^k$ are OPE coefficients, 
$
C_{34}^i(\epsilon)=C^\epsilon\left( (\psi_k^\epsilon)^\ast, \phi_3^\epsilon, \phi_4^\epsilon\right)
$
etc.\ for an appropriate basis $\left\{\psi_i^\epsilon\right\}$ of $\left(\HHH^\WWW\right)^\epsilon$ for some
W-algebra $\WWW$,
and $f_i$ and $\bar{f}_i$ are the corresponding
conformal blocks. Since the family is convergent as $\epsilon\rightarrow0$ and according to
our notion of convergence, 
this factorisation
holds with the same summation in an open neighbourhood of $\epsilon=0$, to which we will restrict 
our discussion in the following.
In the limit $\epsilon\rightarrow 0$, the 
conformal blocks behave as $f_i^\epsilon(z), \bar f_i^\epsilon(\bar z)\rightarrow 1$. 
This fact follows from an analysis of $F^\epsilon_{1234}(z,\ol z)$ under the
conformal transformation $z\mapsto -{1\over z}$ and is the main step in the proof of 
commutativity of $\AAA^\infty$, see \cite[Prop.\ 2.2.4]{rowe03}.
The analyticity assumption implies  
$$
F_{1234}^\epsilon(z,\bar{z})= \sum_i C_{34}^i(0)
C_{2i}^1(0)+\epsilon \eta(z,\bar z) + O(\epsilon^2)\,,
$$
with
$$
\eta(z,\bar z)=\sum_i \left( \left(C_{34}^i
C_{2i}^1\right)^\prime +
C_{34}^i C_{2i}^1 \left(f_i(z)\right)^\prime +
C_{34}^i C_{2i}^1 \left(\bar f_i(\bar z)\right)^\prime\right)\Big|_{\epsilon=0}\,,
$$
where $\left(\cdot\right)^\prime =
{\partial\over\partial\epsilon}\left(\cdot\right)$.
In particular $\bar\partial\partial\eta(z,\bar z)=0$, \ie
$\partial\eta$ is a meromorphic function on $\ol\C$ with singularities in
$\{0,1,\infty\}$. Hence the sum of residues of $\partial\eta$ in the
singularities has to vanish:
\beq\label{res}
\sum_{z\in\{0,1,\infty\}} \Res_z \partial\eta(z) = 0\,.
\eeq
The residues can be easily determined from the OPE-expansion of the
four point functions around the singular points $0,1,\infty$
$$
F_{1234}(z,\bar z)=\left\{\begin{array}{ll}
\sum_\mu C_{34}^\mu C_{2\mu}^1z^{h_\mu-h_3-h_4}\bar{z}^{\bar h_\mu-\bar h_3-\bar
  h_4} & z\approx 0\\
\sum_\mu C_{23}^\mu C_{\mu4}^1(1-z)^{h_\mu-h_2-h_3}(1-\bar{z})^{\bar h_\mu-\bar h_2-\bar
  h_3} & z\approx 1\\
\sum_\mu C_{31}^\mu C_{2\mu}^4z^{h_1-h_3-h_\mu}\bar{z}^{\bar h_1-\bar h_3-\bar
  h_\mu} & {1\over z}\approx 0 \end{array}\right. .
$$
Hence, the singular part of $\partial\eta$ is given by
$$
\left(\partial\eta(z)\right)_{\rm sing}=
\left\{\begin{array}{ll}
{1\over z} \sum_i C_{34}^iC_{2i}^1\left({h_i^\prime-h_3^\prime-h_4^\prime}\right)\big|_{\epsilon=0}& z\approx 0\\[3pt]
{1\over 1-z}\sum_i
C_{23}^iC_{i4}^1\left({h_i^\prime-h_2^\prime-h_3^\prime}\right)\big|_{\epsilon=0} &
z\approx 1\\[3pt]
z\sum_i
C_{31}^iC_{2i}^4\left({h_1^\prime-h_3^\prime-h_i^\prime}\right)\big|_{\epsilon=0} & {1\over z}\approx 0 \end{array}\right.\,,
$$
where the sum reduces to those $i$ with $h_i\rightarrow 0$, because for all other $i$ $C_{ab}^i(0)=0$, $a,b\in\{1,2,3,4\}$.
Since $\sum_i C_{34}^iC_{2i}^1 |_{\epsilon=0}=\sum_i
C_{23}^iC_{i4}^1|_{\epsilon=0}=\sum_i C_{31}^iC_{2i}^4|_{\epsilon=0}$, 
which is also needed to show commutativity of $\AAA^\infty$, \cf
\cite[Lemma 2.2.3]{rowe03},
the residue formula \eq{res} turns into
\beqn\label{resformula}
&&
\smash{\sum_i} \left(\left(C_{34}^iC_{2i}^1 + C_{23}^i C_{i4}^1+  C_{31}^iC_{2i}^4\right)h_i^\prime
\right.\\
&&\qquad\qquad\qquad\qquad\left.- 
C_{34}^iC_{2i}^1\left(h_1^\prime+h_2^\prime+h_3^\prime+h_4^\prime\right)
  \right)\Big|_{\epsilon=0}=0\,.\nonumber
\eeqn
Because of
analyticity, $h_i=\epsilon h_i^\prime |_{\epsilon=0} + O(\epsilon^2)$ 
with $h_i^\prime |_{\epsilon=0}$ finite. Therefore, $L_0$ can be rescaled by $1\over\epsilon$ 
in the limit $\epsilon\rightarrow 0$ (the rescaling function is $r(\epsilon)=\epsilon$), and 
$h_i^\prime |_{\epsilon=0}+\ol h_i^\prime |_{\epsilon=0}$ are the eigenvalues of a corresponding
rescaled operator $H^\infty$ on
$\phi_i\in\H^\infty$. Equation \req{resformula}
together with
$\sum_iC_{31}^iC_{2i}^4|_{\epsilon=0}=\sum_iC_{24}^iC_{3i}^1|_{\epsilon=0}$
implies the $7$-term identity
\begin{eqnarray*}
&&
\hspace*{-2em}
\phi_2 H^\infty(\phi_3\phi_4)
+ H^\infty(\phi_2\phi_3)\phi_4
+ H^\infty(\phi_4\phi_2)\phi_3\\
&&
\hspace*{-1em}
-H^\infty(\phi_2)\phi_3\phi_4
-\phi_2 H^\infty(\phi_3)\phi_4
-\phi_2\phi_3 H^\infty(\phi_4)
-H^\infty(\phi_2\phi_3\phi_4)
\;=\;0,
\end{eqnarray*}
which holds for all $\phi_2,\phi_3,\phi_4\in\H^\infty$
and by misuse of notation which identifies $\phi_i$ with
$A_{\phi_i}\in\AAA^\infty$ via $\phi_i=A_{\phi_i}(\Omega)$.
\section{Conclusions and outlook}\el{conc}
In this note, we have summarised the basic ideas behind our work \cite{rowe03},
which  establishes a new intrinsic notion of
limiting processes for unitary CFTs and of degenerate limits.
We show how our
definition allows to associate geometric interpretations to 
degenerate limits, by applying techniques from noncommutative geometry,
as in \cite{frga93}. Our definitions ensure that
the degenerate limits carry a structure
similar to that predicted by Kontsevich/Soibelman 
in the context of mirror symmetry on tori \cite{koso00}.

The note also contains some new results. For the large level limit of
unitary Virasoro minimal models of A-type we had already obtained a 
sensible geometric interpretation in \cite{rowe03}. We are now able to 
explain how this result is in accord with predictions in \cite{frga93}, obtained by means of the
coset construction for these models along with general
techniques developed for gauged WZW models. Moreover, following ideas
of Maxim Kontsevich, we include a discussion of the so-called seven-term
identity which we show to be satisfied by
the rescaled energy operator in any degenerate limit of CFTs. As we discuss,
this identity can be viewed
as an analogue of the Leibniz rule, characterising second order differential
operators without constant term and with symmetric symbol. 
This allows us to interpret the energy operator in the limit 
as a generalised Laplacian, a property which had to be assumed separately in 
\cite{rowe03}.

There are many interesting open questions which should be addressed
in the context of degeneration phenomena in CFT.
One promising route is the establishment of a larger class of examples
for which our techniques can be proved to work explicitly. E.g.\
for the D-series of unitary Virasoro minimal models one obtains the 
expected result, as we shall show elsewhere. Certainly, 
a generalisation of our large level limits
to arbitrary WZW models and their cosets should be possible and worthwhile.
Another good direction is the inclusion of supersymmetry into the picture.
The expected structure of degenerate limits should include a topological
space with a Riemannian metric, a dilaton, and a complex structure along
with appropriate Dirac operators. Thus the inclusion of supersymmetry is
expected to yield two advantages: On the geometric
side, we expect to find precisely those data which enter in Connes' noncommutative
geometry. On the CFT side, supersymmetry tends to dramatically
simplify the structure of the relevant representation theory. 
These ideas shall be addressed in future work.
\section*{Acknowledgments}
It is a pleasure to thank Maxim Kontsevich and Yan Soibelman 
for helpful comments or discussions. 

D.\ R.\ was supported by a DFG research fellowship and partially by DOE grant DE-FG02-96ER40959.
K.\ W.\ was supported by DFG grant  WE 4340/1-1 
under Schwerpunktprogramm 1154.
%
%
%%%%%%%%%%%%%%%%%%%%%%%%%%%%%%%%%%%%%%%%%%%%%%%%%%%%%%%%%%%%%

\end{document}